\documentclass[zpreprint,zbstdefault,twoside]{zeus_paper}
\usepackage[english]{babel}

\newcommand{\ZcoosysB}{%
The ZEUS coordinate system is a right-handed Cartesian system, with the $Z$
axis pointing in the proton beam direction, referred to as the ``forward
direction'', and the $X$ axis pointing left towards the centre of HERA.
The coordinate origin is at the nominal interaction point.\xspace}

\newcommand{\ZcoosysfnB}{\footnote{\ZcoosysB}}

\newcommand{\Zctddesc}[1]{%
Charged particles are tracked in the central tracking detector (CTD)~\citeCTD,
which operates in a magnetic field of $1.43\Tesla$ provided by a thin 
superconducting solenoid. The CTD consists of 72~cylindrical drift chamber 
layers, organised in nine superlayers covering the polar-angle#1 region 
\mbox{$15^\circ<\theta<164^\circ$}. The transverse-momentum resolution for
full-length tracks is $\sigma(p_T)/p_T=0.0058p_T\oplus0.0065\oplus0.0014/p_T$,
with $p_T$ in $\Gev$.}
\newcommand{\Zcaldesc}{%
The high-resolution uranium--scintillator calorimeter (CAL)~\citeCAL consists 
of three parts: the forward (FCAL), the barrel (BCAL) and the rear (RCAL)
calorimeters. Each part is subdivided transversely into towers and
longitudinally into one electromagnetic section (EMC) and either one (in RCAL)
or two (in BCAL and FCAL) hadronic sections (HAC). The smallest subdivision of
the calorimeter is called a cell.  The CAL energy resolutions, as measured under
test-beam conditions, are $\sigma(E)/E=0.18/\sqrt{E}$ for electrons and
$\sigma(E)/E=0.35/\sqrt{E}$ for hadrons, with $E$ in $\Gev$.}

\chardef\usc=95
\chardef\til=126
\catcode`\@=11 
\DeclareRobustCommand\xdotspace{\futurelet\@let@token\@xdotspace}
\def\@xdotspace{%
  \ifx\@let@token.\else
  \ifx\@let@token\bgroup.\else
  \ifx\@let@token\egroup.\else
  \ifx\@let@token\/.\else
  \ifx\@let@token\ .\else
  \ifx\@let@token~.\else
  \ifx\@let@token!.\else
  \ifx\@let@token,.\else
  \ifx\@let@token:.\else
  \ifx\@let@token;.\else
  \ifx\@let@token?.\else
  \ifx\@let@token/.\else
  \ifx\@let@token'.\else
  \ifx\@let@token).\else
  \ifx\@let@token-.\else
  \ifx\@let@token\@xobeysp.\else
  \ifx\@let@token\space.\else
  \ifx\@let@token\@sptoken.\else
   .\space
   \fi\fi\fi\fi\fi\fi\fi\fi\fi\fi\fi\fi\fi\fi\fi\fi\fi\fi}
\catcode`\@=12 

\newcommand{\stru}[2]{%
   \relax\ifmmode\hbox{\vrule height#1 depth#2 width0pt}%
   \else\vrule height#1 depth#2 width0pt\fi}

\newcommand{\Ronum}[1]{\uppercase\expandafter{\romannumeral#1}}
\newcommand{\ronum}[1]{\expandafter{\romannumeral#1}}
\DeclareRobustCommand{\LaTeXZ}{%
  \LaTeX\kern-.05em4\kern-.1em
  {\raisebox{-0.2ex}{$\scriptstyle\text{ZEUS}$}}\xspace}

\DeclareMathAlphabet{\mathbf}{OT1}{cmr}{bx}{sl}
\newcommand{\eVdist}{\kern-0.06667em}

\newcommand{\Gev}{{\text{Ge}\eVdist\text{V\/}}}

\newcommand{\gev}{{\,\text{Ge}\eVdist\text{V\/}}}

\newcommand{\Tesla}{\,\text{T}}

\newcommand{\slashfrac}[2]{%
  \raisebox{0.5ex}{\ensuremath #1}\kern-0.12em/\kern-0.08em
  \raisebox{-.8ex}{\ensuremath #2}}

\newcommand{\sqr}[3]{%
    {\vcenter{\hrule height.#3ex\hbox{\vrule width.#2ex height#1ex
     \kern#1ex\vrule width.#3ex}\hrule height.#2ex}}}

\newcommand{\widebar}[1]{%
   \mkern1.5mu\overline{\mkern-1.5mu#1\mkern-1.mu}\mkern1.mu}
\catcode`\@=11 
\newcommand{\parenbar}{\mathpalette\p@renb@r}
\def\p@renb@r#1#2{\vbox{%
  \ifx#1\scriptscriptstyle \dimen@.7em\dimen@ii.2em\else
  \ifx#1\scriptstyle \dimen@.8em\dimen@ii.25em\else
  \dimen@1em\dimen@ii.4em\fi\fi \offinterlineskip
  \ialign{\hfill##\hfill\cr
    \vbox{\hrule width\dimen@ii}\cr
    \noalign{\vskip-.3ex}%
    \hbox to\dimen@{$\mathchar300\hfil\mathchar301$}\cr
    \noalign{\vskip-.3ex}%
    $#1#2$\cr}}}
\catcode`\@=12 

\newcommand{\IP}{{\rm I$\kern-0.01667em$P}\xspace}

\mathchardef\qsm=63
\mathchardef\pls=43
\mathchardef\mns=512
\mathchardef\plm=518
\mathchardef\eql=61
\mathchardef\smallleft=300
\mathchardef\smallright=301
\mathchardef\les=316
\mathchardef\gre=318
\mathchardef\leq=532
\mathchardef\grq=533

\catcode`\@=11 
\newcounter{pict@width}
\newcounter{pict@height}
\newlength{\pict@scale}
\newcommand{\psfigadd}[4]{%
\setcounter{pict@width}{1*\ratio{#2+\pict@scale/2}{\pict@scale}}
\setcounter{pict@height}{1*\ratio{#3+\pict@scale/2}{\pict@scale}}
\setlength{\unitlength}{\pict@scale}
\hbox to #2{\hspace{-\fill}\begin{picture}(\thepict@width,\thepict@height)
\put(0,0){\psfig{figure=#1,width=#2,height=#3,clip=}}
\SetScale{0.283466457}
\SetWidth{1.763889}
{#4}
\end{picture}}
}
\newcounter{pict@widthfst}
\newcounter{pict@widthscd}
\newcounter{pict@widthtot}
\newcommand{\psfigaddtwo}[7]{%
\setcounter{pict@widthfst}{1*\ratio{#2+\pict@scale/2}{\pict@scale}}
\setcounter{pict@widthscd}{1*\ratio{#2+#4+\pict@scale/2}{\pict@scale}}
\setcounter{pict@widthtot}{1*\ratio{#2+#4+#6+\pict@scale/2}{\pict@scale}}
\setcounter{pict@height}{1*\ratio{#3+\pict@scale/2}{\pict@scale}}
\setlength{\unitlength}{\pict@scale}
\hbox{\hspace{-\fill}\begin{picture}(\thepict@widthtot,\thepict@height)
\put(0,0){\psfig{figure=#1,width=#2,height=#3,clip=}}
\put(\thepict@widthscd,0){\psfig{figure=#5,width=#6,height=#3,clip=}}
\SetScale{0.283466457}
\SetWidth{1.763889}
{#7}
\end{picture}}
}
\newcommand{\psfigror}[4]{%
\setcounter{pict@width}{1*\ratio{#2+\pict@scale/2}{\pict@scale}}
\setcounter{pict@height}{1*\ratio{#3+\pict@scale/2}{\pict@scale}}
\setlength{\unitlength}{\pict@scale}
\hbox{\begin{picture}(\thepict@width,\thepict@height)
\put(0,\thepict@height){\psfig{figure=#1,width=#3,height=#2,clip=,angle=270}}
\SetScale{0.283466457}
\SetWidth{1.763889}
{#4}
\end{picture}}
}
\newcommand{\psfigrol}[4]{%
\setcounter{pict@width}{1*\ratio{#2+\pict@scale/2}{\pict@scale}}
\setcounter{pict@height}{1*\ratio{#3+\pict@scale/2}{\pict@scale}}
\setlength{\unitlength}{\pict@scale}
\hbox{\begin{picture}(\thepict@width,\thepict@height)
\put(0,0){\psfig{figure=#1,width=#3,height=#2,clip=,angle=90}}
\SetScale{0.283466457}
\SetWidth{1.763889}
{#4}
\end{picture}}
}
\catcode`\@=12 
\newlength\listtextwidth

\catcode`\@=11 
\newlength{\@tabfninsert}
\newlength{\@tabfnwidth}
\newcommand{\tabfootnote}[2]{%
  \setlength{\@tabfninsert}{0.8em}
  \setlength{\@tabfnwidth}{\textwidth}
  \addtolength{\@tabfnwidth}{-\@tabfninsert}
  \addtolength{\@tabfnwidth}{-0.4em}
  \noindent\makebox[\@tabfninsert][r]{\footnotesize$^{#1}$\hfil}\hfill%
  \parbox[t]{\@tabfnwidth}{\footnotesize #2\hfill}}
\catcode`\@=12 

\def\xgo{x_\gamma^{\rm obs}}
\def\kt{{\rm k}_{\rm T}}

\def\et2{\overline{E}_T^2}
\def\lesssim{\;\scriptstyle\stackrel{<}{\sim}\displaystyle}
\def\gtrsim{\;\scriptstyle\stackrel{>}{\sim}\displaystyle}

\def\q2{Q^2}
\def\etm2{\widebar{E_T^2}}

\def\g2{GeV$^2$}

\def\kt{k_T}

\def\etjet{E_T^{\rm jet}}
\def\etjetd{(E_T^{\rm jet})^2}
\def\etjeta{E_T^{\rm jet1}}
\def\etjetb{E_T^{\rm jet2}}
\def\etajet{\eta^{\rm jet}}

\def\disaster{{\sc Disaster++}}

\def\citeCTD{{\cite{%
nim:a279:290,*npps:b32:181,*nim:a338:254%
}}\xspace}
\def\citeCAL{{\cite{%
nim:a309:77,*nim:a309:101,*nim:a321:356,*nim:a336:23%
}}\xspace}

\begin{document}
\prepnum{DESY 04-053}

\title{
The dependence of dijet production on photon virtuality in $ep$  collisions at HERA
}                                                       
                    
\author{ZEUS Collaboration}
\date{March 26th 2004}

\abstract{
The dependence of dijet production on the virtuality of the exchanged photon,
$Q^2$, has been studied by measuring dijet cross sections in the range
$0 \lesssim Q^2 < 2000 {\;\rm GeV}^2$ with the ZEUS detector at HERA 
using an integrated luminosity of 38.6~pb$^{-1}$.  
Dijet cross sections were measured for jets with
transverse energy $E_T^{\rm jet}>7.5$ and $6.5$~GeV and pseudorapidities in
the photon-proton centre-of-mass frame in the range $-3<\eta^{\rm jet}<0$.
The variable $\xgo$, a measure of the photon momentum
entering the hard process, was used to enhance the sensitivity of the measurement
to the photon structure. The $Q^2$ dependence of the ratio of low- to high-$\xgo$ events was measured.
Next-to-leading-order QCD predictions were found to generally underestimate the low-$\xgo$
contribution relative to that at high $\xgo$.
Monte Carlo models based on leading-logarithmic parton-showers, using a partonic
structure for the photon which falls smoothly with increasing $Q^2$,
provide a qualitative description of the data.
}

\makezeustitle

\begin{center}                                                                                     
{                      \Large  The ZEUS Collaboration              }                               
\end{center}                                                                                       
  S.~Chekanov,                                                                                     
  M.~Derrick,                                                                                      
  J.H.~Loizides$^{   1}$,                                                                          
  S.~Magill,                                                                                       
  S.~Miglioranzi$^{   1}$,                                                                         
  B.~Musgrave,                                                                                     
  J.~Repond,                                                                                       
  R.~Yoshida\\                                                                                     
 {\it Argonne National Laboratory, Argonne, Illinois 60439-4815}, USA~$^{n}$                       
\par \filbreak                                                                                     
  M.C.K.~Mattingly \\                                                                              
 {\it Andrews University, Berrien Springs, Michigan 49104-0380}, USA                               
\par \filbreak                                                                                     
  N.~Pavel \\                                                                                      
  {\it Institut f\"ur Physik der Humboldt-Universit\"at zu Berlin,                                 
           Berlin, Germany}
% ...
\newpage                                                                                           
$^{\    1}$ also affiliated with University College London, London, UK \\                          
% ...
\newpage   
\begin{tabular}[h]{rp{14cm}}                                                                       
$^{a}$ &  supported by the Natural Sciences and Engineering Research                               
          Council of Canada (NSERC) \\                                                             
$^{b}$ &  supported by the German Federal Ministry for Education and                               
          Research (BMBF), under contract numbers HZ1GUA 2, HZ1GUB 0, HZ1PDA 5, HZ1VFA 5\\         
$^{c}$ &  supported by the MINERVA Gesellschaft f\"ur Forschung GmbH, the                          
          Israel Science Foundation, the U.S.-Israel Binational Science                            
          Foundation and the Benozyio Center                                                       
          for High Energy Physics\\                                                                
$^{d}$ &  supported by the German-Israeli Foundation and the Israel Science                        
          Foundation\\                                                                             
$^{e}$ &  supported by the Italian National Institute for Nuclear Physics (INFN) \\                
$^{f}$ &  supported by the Japanese Ministry of Education, Culture,                                
          Sports, Science and Technology (MEXT) and its grants for                                 
          Scientific Research\\                                                                    
$^{g}$ &  supported by the Korean Ministry of Education and Korea Science                          
          and Engineering Foundation\\                                                             
$^{h}$ &  supported by the Netherlands Foundation for Research on Matter (FOM)\\                   
$^{i}$ &  supported by the Polish State Committee for Scientific Research,                         
          grant no. 620/E-77/SPB/DESY/P-03/DZ 117/2003-2005\\                                      
$^{j}$ &  partially supported by the German Federal Ministry for Education                         
          and Research (BMBF)\\                                                                    
$^{k}$ &  supported by RF President grant N 1685.2003.2 for the leading                            
          scientific schools and by the Russian Ministry of Industry, Science                      
          and Technology through its grant for Scientific Research on High                         
          Energy Physics\\                                                                         
$^{l}$ &  supported by the Spanish Ministry of Education and Science                               
          through funds provided by CICYT\\                                                        
$^{m}$ &  supported by the Particle Physics and Astronomy Research Council, UK\\                   
$^{n}$ &  supported by the US Department of Energy\\                                               
$^{o}$ &  supported by the US National Science Foundation\\                                        
$^{p}$ &  supported by the Polish Ministry of Scientific Research and Information                  
          Technology, grant no. 112/E-356/SPUB/DESY/P-03/DZ 116/2003-2005\\                        
$^{q}$ &  supported by the Polish State Committee for Scientific Research,                         
          grant no. 115/E-343/SPUB-M/DESY/P-03/DZ 121/2001-2002, 2 P03B 07022\\                    
\end{tabular}                                                                                      

\pagenumbering{arabic} 
\pagestyle{plain}
\raggedbottom

\section{Introduction}
\label{sec-int}

Interactions involving real or quasi-real photons ($Q^2 \approx 0$, where $Q^2$ is the virtuality of the photon)
are well described by calculations that use a partonic structure for the photon~\cite{np:b281:365,*phr:345:265}.
However, in deep inelastic scattering (DIS), where $Q^2$ is large, the virtual photon
is commonly treated as a point-like object and used as a probe of the partonic structure of
nucleons~\cite{prl:23:930,*prl:23:935,*arevns:22:203}.
In this paper, dijet production is investigated
over a large range of incident photon virtualities, including photoproduction, DIS,
and  the transition region between them.
Both the H1~\cite{epj:c13:397,*h1:2004:virph} and ZEUS~\cite{pl:b479:37} collaborations have previously studied
the transition between photoproduction and DIS by measuring inclusive jet and dijet cross sections
in ep collisions.

Two processes contribute to the jet photoproduction cross section at leading order (LO)
in quantum chromodynamics (QCD)~\cite{owen1,*owen2,*owen3,*owen4,drees1,*drees2,*drees3,*drees4,*drees5,*drees6}:
direct, in which the photon
couples as a point-like particle to quarks in the hard scatter; and resolved, in which the photon acts
as a source of partons. Both processes can lead to two jets in the final state. The $\xgo$ variable, which is the
fraction of the photon momentum participating in the production of the dijet system, is used to separate the two
processes since resolved (direct) processes dominate at low (high) $\xgo$ values~\cite{pl:b322:287,*pl:b348:665}. 

In conventional fixed-order QCD calculations, only point-like photon interactions contribute to jet production in DIS.
However, two scales play a role in the interaction: $Q$ and the jet transverse energy, $E_T^{\rm jet}$.
For high $Q^2$ ($Q^2\gg (E_T^{\rm jet})^2$), QCD predicts that
the photon will behave as a point-like object. For $Q^2\ll (E_T^{\rm jet})^2$,
the photon may have an effective partonic structure, even for relatively large values of $Q^2$,
which is resolved at a scale related to the transverse energy of the jets.
Therefore, resolved processes may contribute significantly to the jet cross section. The ratio of cross sections
evaluated in different $\xgo$ ranges is particularly sensitive to the resolved component. 

In this paper, the validity of the above approaches in photoproduction and DIS is studied by measuring dijet cross sections
differential with respect to $Q^2$, $E_T^{\rm jet1}$ and $\eta^F$, where $E_T^{\rm jet1}$ is the $E_T$ of the
jet in the accepted rapidity range which has the highest transverse energy,
and $\eta^F$ is the pseudorapidity of the most forward jet.
The ratio of low- to high-$\xgo$ components is measured as a function of $Q^2$ in different regions
of $\et2$, where $\et2$ is the square of the average transverse energy of the two jets with
highest transverse energy.

The data used in this analysis correspond to an integrated luminosity six times larger than that used
in the previous ZEUS study~\cite{pl:b479:37}. Next-to-leading-order (NLO) QCD
calculations~\cite{hep-ph-9710244,np:b485:291,np:b467:399,*np:b507:295,np:b507:315} have been compared
to measurements that span a large range of photon virtualities. The predictions of leading-logarithm
parton-shower (PS) Monte Carlo (MC) models are compared to the data in the transition region between
photoproduction and DIS, where current NLO calculations are not applicable.

\section{Experimental set-up}
\label{sec-exp}

The data were collected during the 1996 and 1997 running periods, when
HERA operated with protons of energy $E_p=820$~GeV and 
positrons of energy $E_e=27.5$~GeV, and correspond to an integrated
luminosity of $38.6\pm 0.6$~pb$^{-1}$.

The ZEUS detector is described in detail
elsewhere~\cite{zeus:1993:bluebook}. The most important components
used in the current analysis were
the central tracking detector (CTD),
the uranium-scintillator calorimeter (CAL)
and the beam pipe calorimeter (BPC). 

\Zctddesc\ZcoosysfnB

\Zcaldesc

The BPC~\cite{pl:b407:432} was installed  294 cm from the interaction
point in the positron direction in order to tag scattered positrons at
small angles with respect to the positron beam direction (15-34 mrad).
It measured both the energy and impact position  of the
scattered positron at the BPC surface. The relative energy
resolution of the BPC is $0.17/\sqrt{E}$ and the position resolution
is 0.5 mm.

The luminosity was determined from the rate of the bremsstrahlung process
$ep \rightarrow e \gamma p$, where the photon was measured with a
lead-scintillator
calorimeter~\cite{desy-92-066,*zfp:c63:391,*acpp:b32:2025} at
$Z=-107$~m.

\section{Theoretical framework}
\label{sec-theor}

In photoproduction, perturbative QCD (pQCD) calculations of dijet cross sections can be
written as a convolution of the subprocess cross section with the
parton distribution functions (PDFs) of the photon and proton:
$$ d\sigma_{e p \rightarrow e\, {\rm jet}\,{\rm jet} } = 
 \displaystyle\sum_{a,b} \int_0^1 dy f_{\gamma/e}(y,\mu_R^2) \int_0^1
 dx_{\gamma} 
   f_{a/\gamma}(x_{\gamma},\mu_R^2,\mu^2_{F\gamma})\ \times$$
\begin{flushright}
$\displaystyle\int_0^1 dx_{p} f_{b/p}(x_{p},\mu^2_{Fp}) \;
   d\hat{\sigma}_{ab \rightarrow {\rm jet}\,{\rm jet}}(\mu_R),$
\end{flushright}
where $y$, $x_{\gamma}$ and $x_p$ are the longitudinal momentum
fractions of the almost-real photon emitted by the positron, the parton
$a$ in the photon and the parton $b$ in the proton, respectively.
The function $f_{\gamma/e}$ is the flux of photons from the positron, and
$f_{a/\gamma}$ ($f_{b/p}$) represents the PDF of parton $a$ ($b$) in the photon (proton).
The factorisation scale for the photon (proton) is denoted by $\mu_{F\gamma}$ ($\mu_{Fp}$) and $\mu_R$
represents the renormalisation scale. The subprocess cross
section, $d\hat{\sigma}_{ab\rightarrow {\rm jet}\,{\rm jet}}$,
describes the short-distance structure of the interaction.
For direct processes in the above formula $a$ is replaced by
$\gamma$ and $f_{a/\gamma}(x_{\gamma},\q2,\mu^2_{F\gamma})$ is given by $\delta(1-x_\gamma)$.

In DIS, the photons are virtual ($\gamma^*$) and usually
only direct processes are considered. Effective resolved terms
appear only as higher-order corrections.

In the transition region between DIS and photoproduction,
a virtual-photon structure~\cite{pl:b376:193,*drees7,*theory1,*theory2,*theory3,*theory4,pred1,*pred2,*pred3,grs} may be introduced.
In general, the virtual-photon PDFs $f_{a/\gamma^*}$ contain two terms,
$$f_{a/\gamma^*}(x_{\gamma^*},\q2,\mu^2_{F_{\gamma^*}})=f^{\rm non-pert}_{a/\gamma^*}(x_{\gamma^*},\q2,\mu^2_{F_{\gamma^*}})+ 
f^{\rm pert}_{a/\gamma^*}(x_{\gamma^*},\q2,\mu^2_{F_{\gamma^*}}),$$
the first associated with the non-perturbative hadronic component
($f^{\rm non-pert}$), in which the photon fluctuates into an intermediate
meson-like hadronic state, and the second $f^{\rm pert}$, unique to the
photon, which expresses the coupling of the photon to a high-virtuality $q\bar q$ pair,
calculable in pQCD. Perturbative QCD predicts that the contribution to
the dijet cross section from resolved processes should decrease relative
to the contribution from direct processes as the virtuality of the
photon increases towards $\mu_R$.
The non-perturbative component of the virtual-photon PDFs decreases as
$Q^{-4}$, whereas the perturbative component decreases as $\ln(\mu_R^2/\q2)$.

Two parameterisations of the virtual-photon
PDFs, SaS~\cite{pl:b376:193} and GRS~\cite{grs}, are available.
Both are extrapolations of the real-photon
PDFs to the virtual-photon regime. They differ in the treatment of the
non-perturbative component. In the case of the SaS sets, a fit to a
coherent sum of the lowest-lying vector-meson states $\rho$, $\omega$
and $\phi$ has been performed, whereas, in the case of GRS, the
non-perturbative part has been estimated using the PDFs of the pion.

\section{Cross section definition}

Dijet cross sections  differential in $Q^{2}$, $E_T^{\rm jet1}$ and $\eta^{F}$ were measured.
The ratios of cross sections for low ($<0.75$) to high ($>0.75$) $\xgo$ are presented.
The variable $\xgo$ is defined as
\begin{equation} \xgo = \frac{\sum_{\rm jets} ( E^{\rm jet}-p_Z^{\rm jet} )}{\sum_{\rm hadrons} ( E - p_Z ) }\nonumber\;\;,\end{equation}
where $E^{\rm jet}$ and $p_Z^{\rm jet}$ are the energy and the longitudinal momentum of the jet.
The upper sum runs over the two jets with highest transverse energy and the lower sum runs over
all final state hadrons.

The  cross sections were measured in the range $0 \lesssim Q^{2} < 2000$~GeV$^{2}$ and $0.2 < y < 0.55$.
Jets were reconstructed  with the $\kt$ cluster algorithm~\cite{np:b406:187}
applied in the photon-proton centre-of-mass frame,
in the longitudinally invariant inclusive mode~\cite{pr:d48:3160}. 
At least two jets were required within the pseudorapidity range $-3<\etajet<0$, satisfying $\etjeta > 7.5$~GeV and $\etjetb>6.5$~GeV.

\section{Data selection and jet search}
\label{data_selection}

A three-level trigger was used to select events online~\cite{zeus:1993:bluebook,epj:c1:109}. 
In the third-level trigger the events were
required to have at least two jets with a transverse energy of
$E_T^{\rm jet} > 4$~GeV and a pseudorapidity of $\eta^{\rm jet} < 2.5$ in the laboratory frame.

The sample was separated offline into subsamples corresponding to three
different $\q2$ ranges:

\begin{itemize}
\item DIS sample: events were selected by
  requiring that the outgoing positron was measured in the
  CAL~\cite{nim:a365:508}. The energy of the scattered positron, $E_{e^\prime}$, was
  required to be above 10 GeV, with $1.5 < \q2 <2000$ \g2;
\item BPC sample: events at low $\q2$  were selected by requiring that the
  scattered positron was measured in the BPC. These events were required to have
  $E_{e^\prime} > 12.5$~GeV and $0.1 < \q2 < 0.55$ \g2;
\item Photoproduction sample: events were selected by requiring that the scattered positron
  was not observed in the CAL, implying $Q^2<1$~GeV$^2$
  with a median $Q^2\sim 10^{-3}$~GeV$^2$. A small fraction of this sample (0.6~\%)
  is also contained in the BPC sample.
\end{itemize}

For all three samples, hadronic kinematic variables and jets were
reconstructed using a combination of track and CAL information which
optimises the resolution~\cite{briskin:phd:1998}. 
The selected tracks and CAL clusters are referred to as Energy Flow
Objects (EFOs).

The method reported in a previous publication~\cite{epj:c11:35} was used
to correct the EFOs for energy losses in inactive material in front of
the CAL. The jet-energy-scale uncertainty is within
$\pm 1\%$ for $\etjet > 7.5$~GeV and increases to $\pm 3\%$ for lower $\etjet$ values.

Additional cuts, similar to those used in an earlier
analysis~\cite{pl:b479:37}, were applied offline to all samples:

\begin{itemize}
\item a reconstructed event vertex consistent with the nominal
  interaction position was required, $|Z_{\rm vtx}| < 40$~cm;
\item to suppress the background from events with a misidentified
  positron, the variable $y_e=1-\frac{E_{e^\prime}}{2E_{e^\prime}(1-\cos\theta_{e^\prime})}$
  was required to satisfy $y_e < 0.8$, where $\theta_{e^\prime}$ is the polar angle of the
  scattered positron;
\item for the DIS sample, a fiducial volume cut was applied to the
  positron position ($|X_e|>14$ cm or $|Y_e|>9$ cm, where $X_e$ and
  $Y_e$ are the impact positions of the positron on the face of the CAL) in order
  to avoid the low-acceptance region adjacent to the rear beam pipe;
\item for the BPC sample, the reconstructed impact position on the
  BPC surface was constrained to be within the fiducial-region of the
  BPC~\cite{pl:b407:432};
\item for the photoproduction sample, events with a scattered-positron
  candidate in the CAL were rejected, as in a previous ZEUS analysis~\cite{ej:c23:615};
\item all samples were required to satisfy $0.2 < y_{\rm JB} < 0.55$,
  where $y_{\rm JB} = \sum_i (E_i - E_{Zi}) /2E_e$~\cite{proc:epfacility:1979:391} is an estimator of $y$.
  The sum runs over all EFOs.
  $E_{Zi} = E_i \cos\theta_i$, where $E_i$ is the energy of EFO $i$ with polar angle
  $\theta_i$ with respect to the measured $Z$-vertex of the event.
  The lower cut removes beam-gas events and the upper cut is imposed due
  to the restricted acceptance of the BPC detector.
\end{itemize}

Prior to jet finding, the EFOs were boosted to the photon-proton centre-of-mass frame.
In the DIS and BPC samples the boost
was calculated using the reconstructed momentum of the scattered positron.
In the photoproduction sample $y_{\rm JB}$ was used in performing the boost.

The $\kt$ cluster algorithm was applied to the boosted EFOs in the photon-proton centre-of-mass frame
to reconstruct jets.  At least two jets were required in each event
within the pseudorapidity range $-3<\etajet<0$ and were ordered
according to decreasing $\etjet$. They were further required to
satisfy $\etjeta > 7.5$ GeV and $\etjetb>6.5$ GeV.
After all cuts, the photoproduction/BPC/DIS sample contained 419911/2481/45100 dijet events.
The BPC sample is a subset of the photoproduction sample.

\section{Acceptance corrections}

The programs {\sc Herwig~}5.9~\cite{cpc:67:465} and {\sc Pythia~}6.1~\cite{cpc:135:238}
were used to generate events for resolved and direct processes over the whole $Q^2$ range.
Events were generated using GRV-LO~\cite{pr:d46:1973} for the photon
PDFs and MRSA~\cite{pr:d50:6734} for the proton PDFs. 
To study the dependence of the acceptance corrections
on the choice of photon and proton PDFs, the GRS-LO and
CTEQ5M1~\cite{epj:c12:375} parameterisations were used, respectively. 
In both generators, the partonic processes are simulated using LO matrix
elements, with the inclusion of initial- and final-state parton
showers. Hadronisation is performed using a cluster
model~\cite{np:b238:492} in the case of {\sc Herwig} and the Lund
string model~\cite{prep:97:31}
in the case of {\sc Pythia}. For the measurements presented in this paper,
the {\sc Herwig} and {\sc Pythia} programs were used to
correct the data for acceptance. The corrections provided by {\sc Herwig} were used as default
values and those given by {\sc Pythia} were used to estimate the systematic
uncertainties associated with the treatment of the parton shower and
hadronisation.

All generated events were passed through the ZEUS detector and trigger
simulation programs based on {\sc Geant~}3.13~\cite{tech:cern-dd-ee-84-1}. 
They were reconstructed and analysed by the same program chain as the data. The
jet search was performed using EFOs in the same way as for
the data. The same jet algorithm was also applied to the final-state
particles. The jets found in this way are referred to as hadronic jets.

The acceptance corrections take into account the efficiency of the
trigger, the selection criteria and the purity and efficiency of the
jet reconstruction. The differential dijet cross sections were
obtained by applying bin-by-bin corrections to the measured
distributions. The predictions of the generators {\sc Herwig} and {\sc Pythia} for
the uncorrected distributions were compared to the data for
the above parameterisations of the photon and proton PDFs. The
contributions from direct and resolved processes were added according
to a fit to the uncorrected $\xgo$ distribution in the
data. A good description of the $\etjet$, $\etajet$, $\q2$ and $y$
data distributions was given by both {\sc Herwig} and {\sc Pythia}.

For the photoproduction sample the bin-by-bin correction factor was approximately 1.2.
This increased to approximately 6 for the BPC sample due
to the geometric acceptance of the BPC detector~\cite{pl:b407:432}. 
For $1.5<\q2<4.5$ \g2, the correction factor was approximately 3 due to the
fiducial volume cut (see section \ref{data_selection}). For $\q2>4.5$ \g2, the
bin-by-bin correction factors differed from unity by less than $10\%$.

\section{QCD calculations}

\subsection{NLO calculations}

The NLO QCD calculations of jet production cross sections in DIS used in this analysis are based on the
programs {\sc Disaster++}~\cite{hep-ph-9710244} and {\sc Disent}~\cite{np:b485:291}.
In these programs, the photon is treated as a
point-like probe. Contributions from hadron-like resolved processes are not included.
They use the subtraction method~\cite{subm} and the massless $\widebar{\rm MS}$ renormalisation and factorisation schemes. 
Their predictions agree to within $\pm 3\%$.
In Section~\ref{results_section} only the calculations using {\sc Disaster++} are
compared to the data because this program allows a wider parameter
selection than {\sc Disent}.
In the calculations, the number of flavours was set to
five. The renormalisation and factorisation scales were  set to
$\mu^2=\mu_R^2=\mu_F^2=\q2+(\etjet)^2$ or $\q2$,
and $\alpha_s(\mu_R)$ was calculated at two loops using
$\Lambda^{(5)}_{\widebar{\rm MS}}=226$ MeV corresponding to $\alpha_S(M_Z)=0.118$. 
The CTEQ5M1 sets were used for the proton PDFs.

Many calculations of jet photoproduction at NLO
exist~\cite{np:b467:399,*np:b507:295,np:b507:315,pr:d56:4007,pr:d57:5555,zfp:c76:67,epjdirect:c1:1,epj:c17:413},
all of which agree to within $(5-10)\%$~\cite{epj:c17:413,proc:mc:1998:171}.
The calculations of Frixione and Ridolfi~\cite{np:b467:399,*np:b507:295,np:b507:315} uses the subtraction
method. In this calculation the number of flavours was set to five and the
factorisation and renormalisation scales to $\mu^2=(\etjet)^2$. For the calculation
of  $\alpha_s(\mu_R)$, $\Lambda^{(5)}_{\widebar{\rm MS}}=226$ MeV was used.
For the proton PDFs, the CTEQ5M1 sets were used, and for the real photon PDFs
the GRV and AFG~\cite{zfp:c64:621} parameterisations were used.

Samples of events generated using the {\sc Heracles~}4.6.1~\cite{cpc:69:155,*spi:www:heracles}
MC program with the {\sc Djangoh~}1.1~\cite{proc:hera:1991:23,*spi:www:djangoh11,*cpc:81:381} interface to the
hadronisation programs were used to estimate hadronisation corrections
for the NLO QCD predictions calculated using {\sc Disent} and {\sc Disaster++}.
The QCD cascade is simulated using the colour-dipole
model~\cite{cdm1,*cdm2,*cdm3,*cdm4} including the LO QCD diagrams as
implemented in  {\sc Ariadne~}4.08~\cite{cpc:71:15} or with the MEPS model of
{\sc Lepto~}6.5~\cite{cpc:101:108}. Both {\sc Ariadne} and {\sc Lepto} use the Lund string model~\cite{prep:97:31} for
the hadronisation. For the photoproduction NLO prediction, the {\sc Herwig} and {\sc Pythia} MCs
were used to estimate the hadronisation corrections.

First-order QED radiative effects were also estimated using {\sc Heracles} and found to be $1\%$ or less.
Corrections for these effects have not been applied to the NLO calculations.

The predictions to be compared with the
data were corrected for hadronisation effects  using a bin-by-bin procedure
according to $d\sigma=d\sigma^{\rm NLO}\cdot C_{\rm had}^{-1}$, where
$d\sigma^{\rm NLO}$ is the cross section for partons in the final
state of the NLO calculation. The hadronisation correction factor was
defined as the ratio of the dijet cross sections before and after the
hadronisation process,
$C_{\rm had}=d\sigma_{\rm MC}^{\rm partons}/d\sigma^{\rm hadrons}_{\rm
MC}$. The value of $C_{\rm had}$ was taken as the mean of the ratio
obtained using the predictions of two different generators ({\sc Ariadne} and {\sc Lepto}
for DIS, and {\sc Herwig} and {\sc Pythia} for photoproduction)
and was found to lie between 1.1 (large $Q^2$) and
1.2 (small $Q^2$ and photoproduction).

\subsection{Monte Carlo predictions}

Predictions of {\sc Herwig~}6.4~\cite{jhep:01:010} using CTEQ5L for the proton PDFs and
SaS2D for the photon PDFs were generated using parameters
tuned~\cite{cp:153:2003} to many previous HERA and LEP measurements.
In the SaS2D parameterisation the structure
of the virtual photon is suppressed with increasing $Q^2$.
Predictions were also generated with this suppression switched off.

\section{Systematic uncertainties}
\subsection{Experimental uncertainties}
A detailed study of the sources contributing to the systematic
uncertainties of the measurements was performed. This study includes
(a typical contribution to the uncertainty in the cross section
for each item is indicated in parentheses):

\begin{itemize}
\item using the {\sc Pythia} generator to evaluate the acceptance corrections to
  the observed dijet distributions ($+ 6 \%$);
\item using different parameterisations of the photon (GRV-LO and GRS)
  and proton (MRSA and CTEQ5M1) PDFs for the generation of the {\sc Herwig} MC
  samples ($\pm 2\%$);
\item varying the $\etjet$ cut by the resolution ($\pm 8\%$);
\item varying the other selection cuts by their respective resolution ($<\pm 2\%$)
\item adding the contributions from direct and resolved processes
  according to the default cross sections as predicted by {\sc Herwig}
  ($-3 \%$).
\end{itemize}

All the above systematic uncertainties were added in quadrature.
The effect of the uncertainty in the absolute energy scale of the jets
on the dijet cross sections was approximately $\pm 9 \%$ at low $\q2$, decreasing
to $\pm 6\%$ at high $\q2$. 
This uncertainty is highly correlated and is shown separately in the
figures. In addition, there is an overall normalisation uncertainty
from the luminosity determination of $1.6 \%$, which is not shown
in the figures.

\subsection{Theoretical uncertainties}
The NLO QCD predictions for the dijet cross sections are affected by
the following theoretical uncertainties (typical values for the
uncertainties are quoted):
\begin{itemize}
\item uncertainties due to terms beyond NLO, estimated by varying
  $\mu$ by factors 2 and 0.5 ($20\%$ at low $\q2$ and $7\%$ at
  high $\q2$, in the case of $\mu^2=\q2+\etjetd$);
\item uncertainties in the hadronisation corrections, estimated as
  half the spread between $C_{\rm had}$ values obtained using the
  {\sc Herwig}, {\sc Pythia}, {\sc Lepto} and {\sc Ariadne} models ($2-3\%$);
\item uncertainties on the calculations due to $\alpha_S$ and the proton PDFs,
  estimated by using the MRST sets of parameterisations ($5 \%$).
  These uncertainties were cross-checked using an alternative method~\cite{pl:b560:7},
  which uses the covariance matrix of the fitted PDF parameters and derivatives as a function
  of $x$.
\end{itemize}

The above theoretical uncertainties were added in quadrature to give
the total uncertainty on the predictions. 

\section{Results}
\label{results_section}

\subsection{Single-differential dijet cross sections}

\begin{figure}
\centerline{\epsfig{file=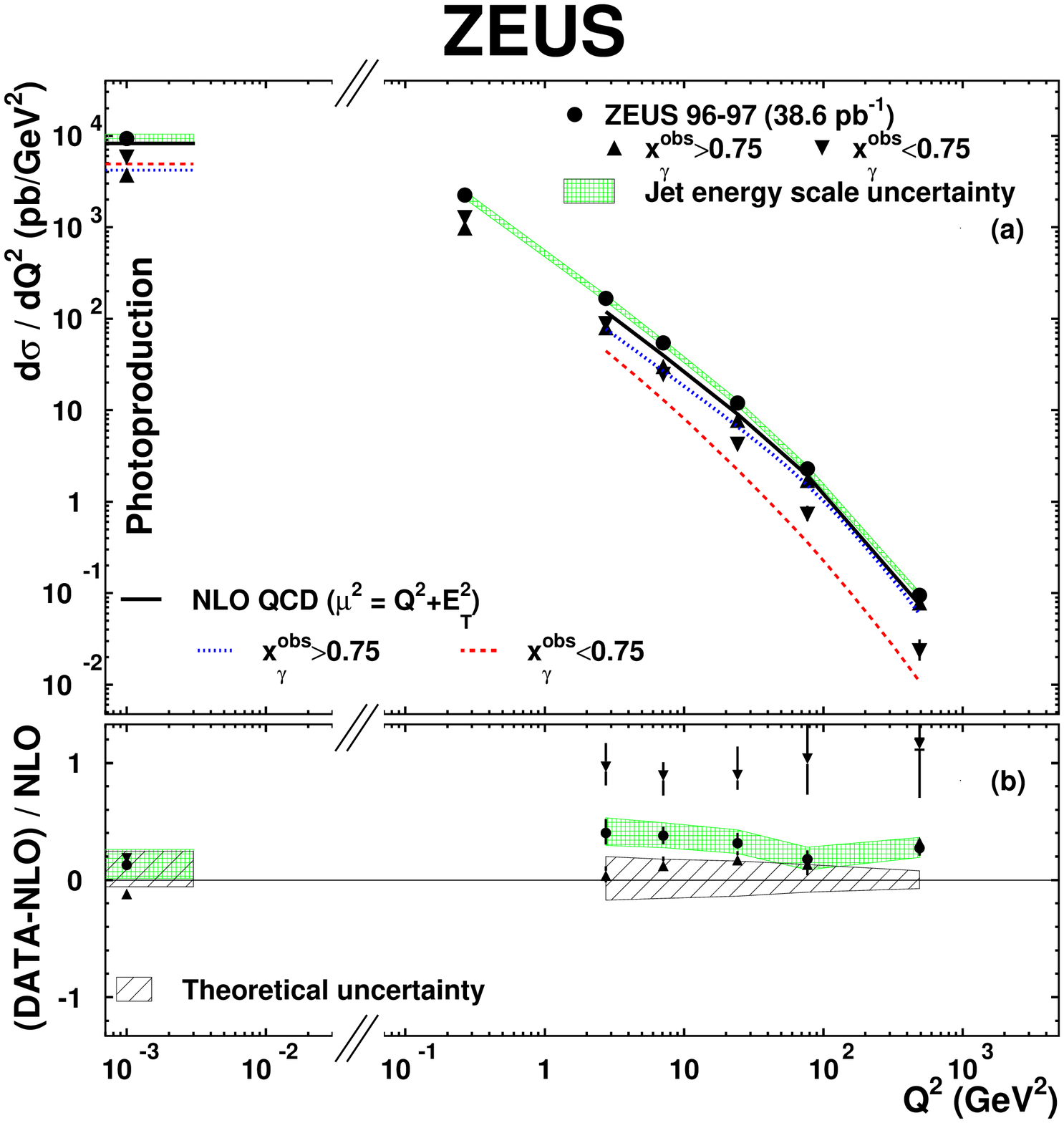,width=12.5cm,clip=}}
\caption{
(a) Measured dijet cross sections $d \sigma / d Q^2$ for $\xgo \geq 0.75$ (upwards triangles)
$d \sigma / d Q^2$ for $\xgo < 0.75$ (downwards triangles) and $d \sigma / d Q^2$ for the entire $\xgo$ 
region (black dots).
The inner vertical bars represent the statistical uncertainties of the data, and the outer bars show
the statistical and systematic uncertainties added in quadrature, except for that associated
with the uncertainty in the absolute energy scale of the jets (shaded band).
The NLO QCD calculations of {\sc Disaster++} ($\mu^2=\q2+\etjetd$)
and of Frixione and Ridolfi ($\mu^2=\etjetd$) for the photoproduction region are shown for each of the cross sections.
(b) Relative difference of the measured dijet cross section $d \sigma / d Q^2$ to the {\sc Disaster++} ($\mu^2=\q2+\etjetd$)
and of Frixione and Ridolfi ($\mu^2=\etjetd$) calculations. The hatched band shows the theoretical uncertainty of the calculations
(see text).}
\label{virtualgamma_fig1}
\end{figure}

Figure~\ref{virtualgamma_fig1} shows the differential dijet cross section, $d\sigma/dQ^2$, for
$E_T^{\rm jet1}>7.5$~GeV, $E_T^{\rm jet2}>6.5$~GeV, $-3<\eta^{\rm jet}<0$, $0.2<y<0.55$ and
$0.1<Q^2<2000$~GeV$^2$ together with the photoproduction measurement. The cross section
split in the  direct-enhanced region ($\xgo \geq 0.75$) and the resolved-enhanced region ($\xgo < 0.75$)
is also shown. All the cross sections are given in the tables~\ref{table-dsdq2}, \ref{table-dsdq2-lox}, and~\ref{table-dsdq2-hix}.

The measurements cover a wide range in $Q^2$,
including the transition region from photoproduction to DIS. The measured cross sections fall by about
five orders of magnitude over this $Q^2$ range. The cross section for $\xgo< 0.75$ falls more rapidly than that
for $\xgo > 0.75$. Even though the dijet cross section is dominated by interactions with $\xgo > 0.75$ for $Q^2\gtrsim
10$~GeV$^2$, there is a contribution of approximately $24\%$ from low-$\xgo$ events with $Q^2 \simeq 500$~GeV$^2$.

\begin{figure}
\centerline{\epsfig{file=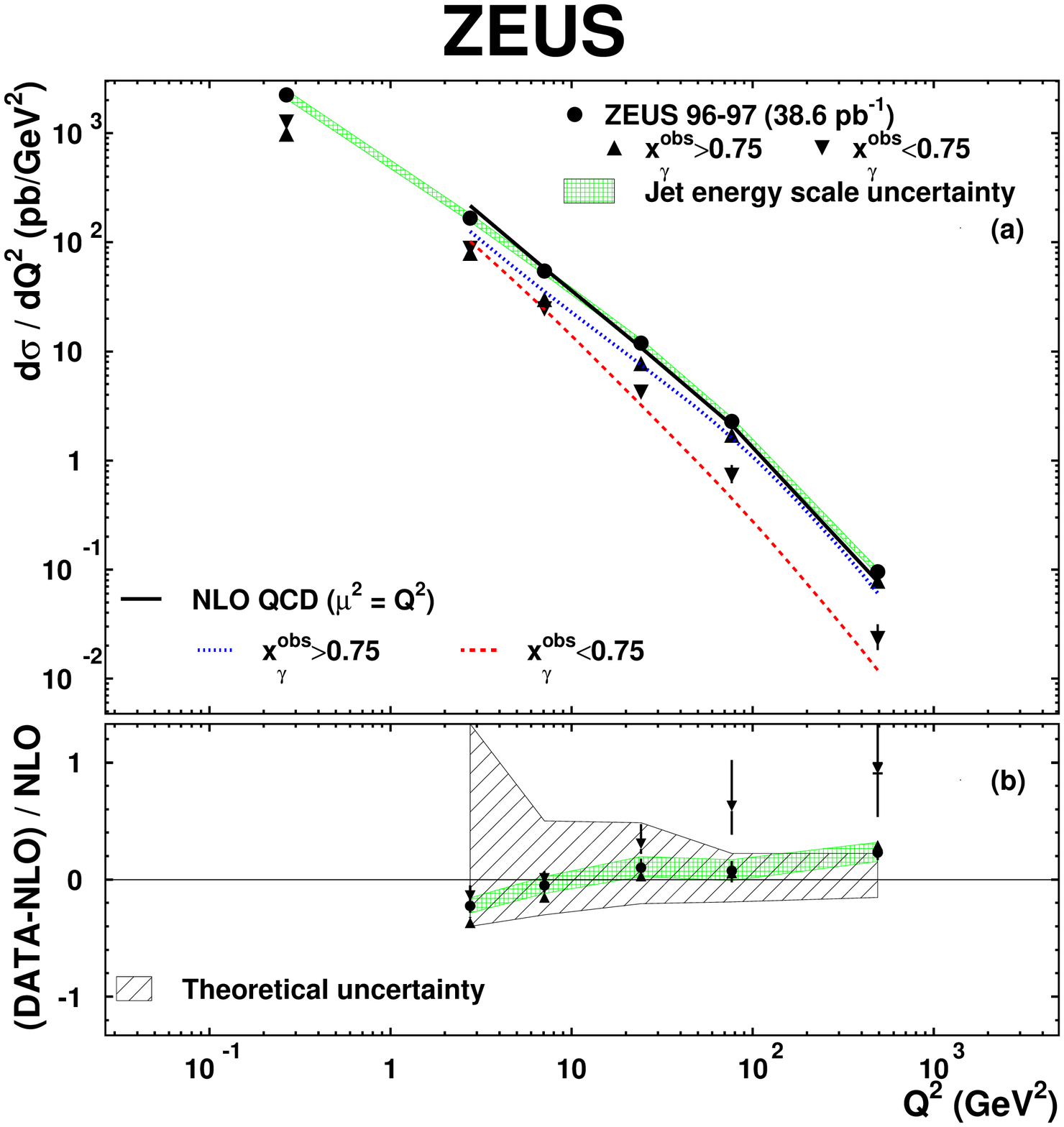,width=12.5cm,clip=}}
\caption{
(a) Measured dijet cross sections $d \sigma / d Q^2$ for $\xgo \geq 0.75$ (upwards triangles)
$d \sigma / d Q^2$ for $\xgo < 0.75$ (downwards triangles) and $d \sigma / d Q^2$ for the entire $\xgo$
region (black dots).
The inner vertical bars represent the statistical uncertainties of the data, and the outer bars show
the statistical and systematic uncertainties added in quadrature, except for that associated
with the uncertainty in the absolute energy scale of the jets (shaded band).
The NLO QCD calculations of {\sc Disaster++} with $\mu^2=\q2$ are shown for each of the cross sections.
(b) Relative difference of the measured dijet cross section $d \sigma / d Q^2$ to the \disaster calculation with
$\mu^2=\q2$. The hatched band shows the theoretical uncertainty of the calculation.}
\label{virtualgamma_fig2}
\end{figure}

The NLO QCD calculations are compared to the measured $d\sigma/dQ^2$
in Figs.~\ref{virtualgamma_fig1} and \ref{virtualgamma_fig2}.  The
prediction\footnote{The two lowest $Q^2$ bins are outside the range of
applicability of the {\sc Disaster++} program.} with
$\mu^2=\q2+\etjetd$, shown in Fig.~\ref{virtualgamma_fig1}, describes
the shape of the measured total dijet cross section but underestimates
its magnitude by approximately $30\%$.  The renormalisation scale
uncertainty was evaluated also for the low- and high-$\xgo$ cross
sections. For the high-$\xgo$ cross section this uncertainty was
similar to that on the total cross section. In the case of the
low-$\xgo$ cross section, the uncertainty was almost constant at
around $\pm$30\%. Taking these uncertainties into account, the
measured cross section for $\xgo>0.75$ is reasonably well described by
the calculation shown in Fig.~\ref{virtualgamma_fig1}a for all
$\q2$. However, the prediction dramatically underestimates the
measured cross section for $\xgo < 0.75$.

The prediction with $\mu^2=\q2$ is shown in
Fig.~\ref{virtualgamma_fig2}. It has a much larger renormalisation-scale
uncertainty than the prediction using $\mu^2=\q2+\etjetd$, and
within this uncertainty it is consistent with the data.

A possible explanation of the disagreement, and for the large uncertainties in
the prediction at low-$\xgo$ values, is that effects arising from the structure
of the photon are expected in this region, whereas the contribution
predicted by {\sc Disaster++} comes only from large-angle
particle-emission diagrams included in the NLO corrections to the
dijet cross section.

In photoproduction, the low-$\xgo$ component of the data becomes
dominant.  The photoproduction measurement is well described by the
photoproduction NLO prediction, using the GRV photon PDF.

\subsection{Double-differential dijet cross sections}

\begin{figure}
\centerline{\epsfig{file=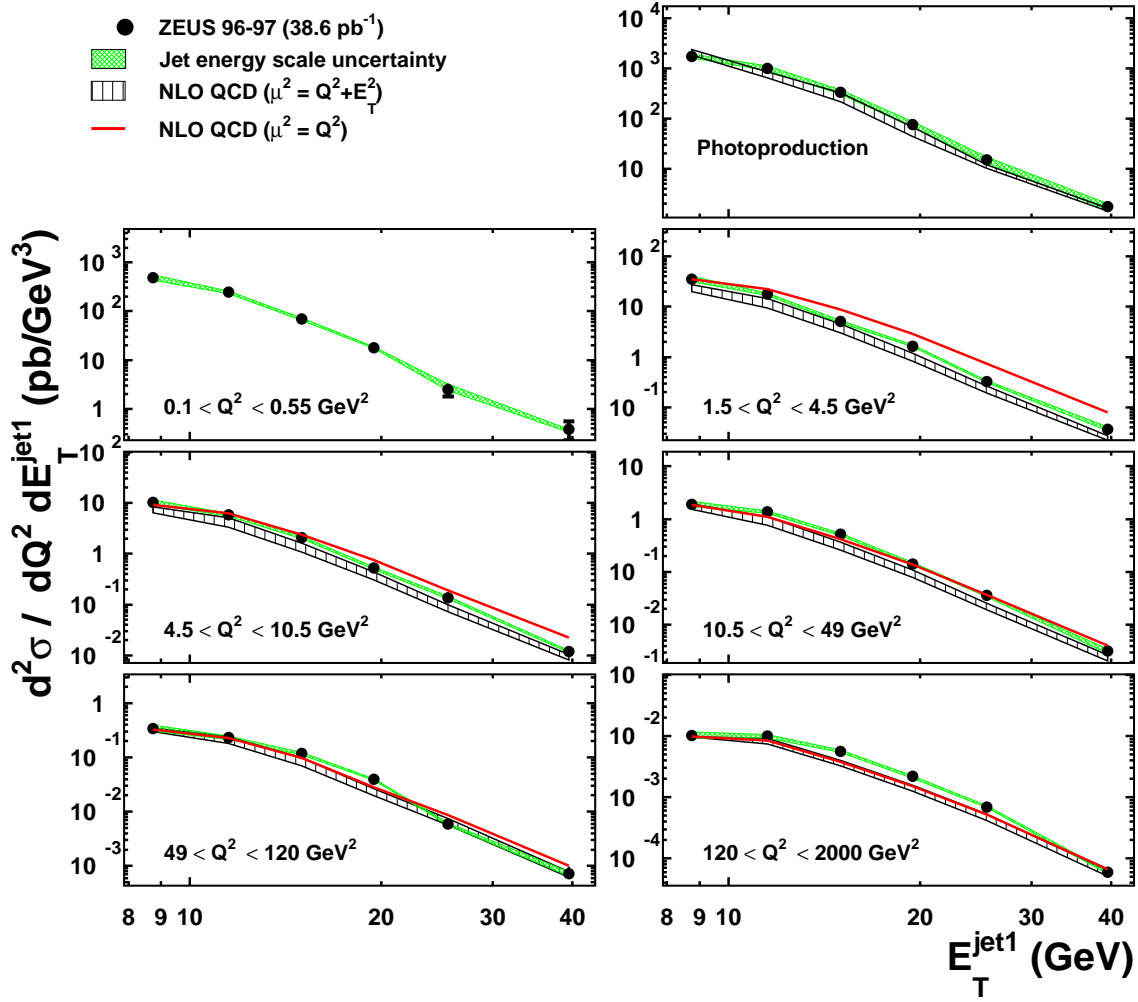,width=15cm,clip=}}
\caption{
Measured dijet cross section $d^2 \sigma / d Q^2 d E_T^{\rm jet1}$ (dots).
Also shown are the NLO QCD calculations of {\sc Disaster++} with  $\mu^2=\q2+\etjetd$
and $\mu^2=\q2$, and those of Frixione and Ridolfi for the photoproduction region.
The scale uncertainty for the NLO calculation with $\mu^2=\q2$ is not shown.
Other details are as in Fig.~1.}
\label{virtualgamma_fig3}
\end{figure}

The dijet cross section, $d^2\sigma/dQ^2dE_T^{\rm jet1}$, as a function of $E_T^{\rm jet1}$ in
different $Q^2$ ranges is shown in Fig.~\ref{virtualgamma_fig3}
and given in the Tables~\ref{table-q2etl-1} and~\ref{table-q2etl-2}.
The measurements extend up to transverse
energies of approximately 40~GeV. The $E_T^{\rm jet1}$ distribution
falls less steeply as $Q^2$ increases.  Figure~\ref{virtualgamma_fig3}
also shows the NLO QCD predictions. The NLO calculation for
photoproduction using GRV for the photon PDFs gives a good description
of the $E_T^{\rm jet1}$ cross section. At higher $Q^2$, the
calculation using $\mu^2=\q2+\etjetd$ is in agreement with the data
for the lowest and highest jet transverse energies, but lies below the
data for intermediate $E_T^{\rm jet1}$ values. The prediction with
$\mu^2 = Q^2$ again agrees with the data, within the large theoretical
uncertainties (not shown).

\begin{figure}
\centerline{\epsfig{file=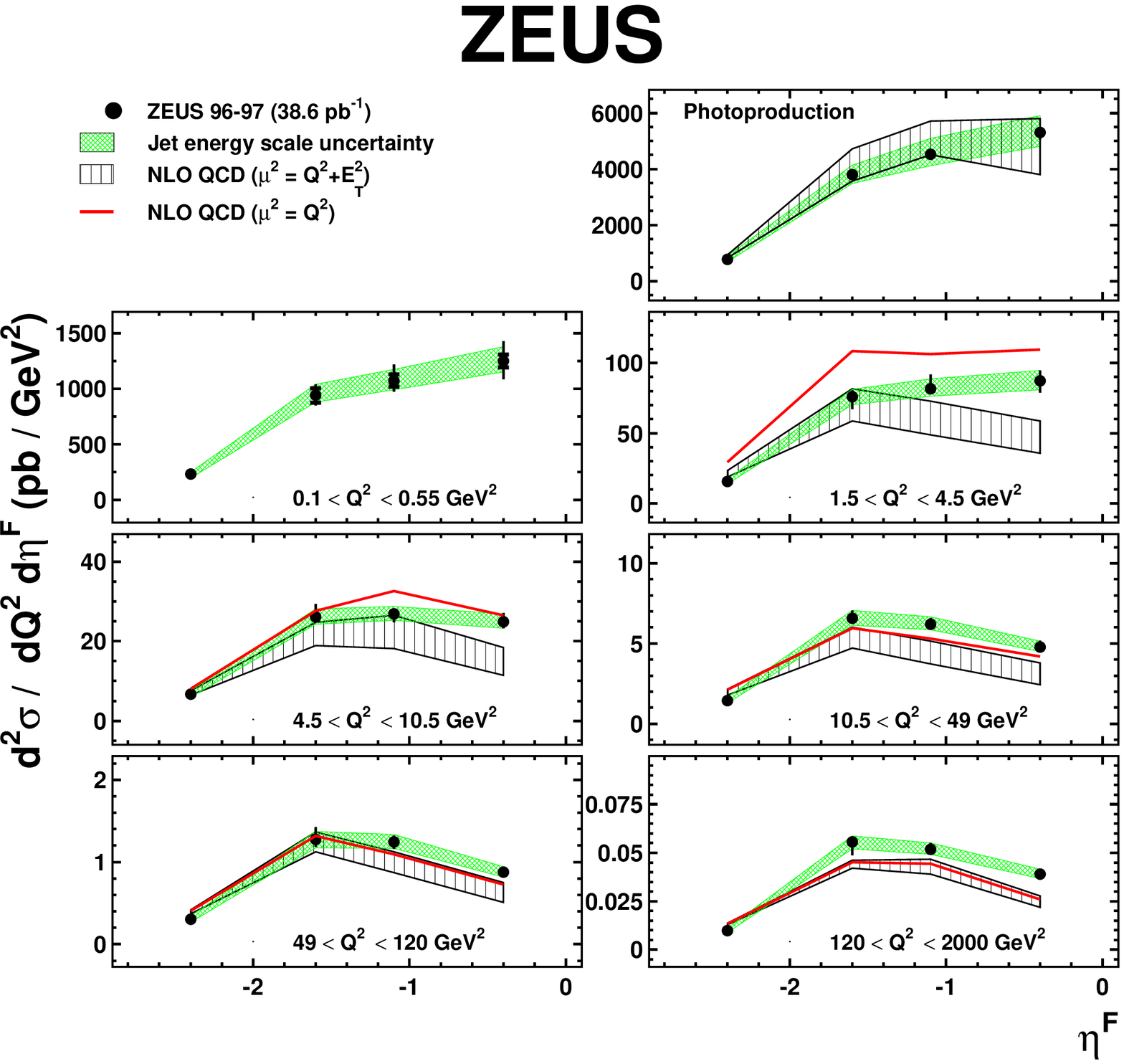,width=15cm,clip=}}
\caption{
Measured dijet cross section $d \sigma / d Q^2 d  \eta^F$ (black
dots).
The NLO QCD calculations of {\sc Disaster++} with  $\mu^2=\q2+\etjetd$
and $\mu^2=\q2$ as well as Frixione and Ridolfi for
the photoproduction region are also shown.
The scale uncertainty for the NLO calculation with $\mu^2=\q2$ is not shown.
Other details are as in Fig.~1.}
\label{virtualgamma_fig4}
\end{figure}

The differential cross-section $d^2\sigma/dQ^2d\eta^F$ as a function
of $\eta^F$ is shown in Fig.~\ref{virtualgamma_fig4} for different
ranges of $Q^2$
and given in the Tables~\ref{table-q2eta-1} and~\ref{table-q2eta-2}.
The cross section as a function of $\eta^F$ is more
sensitive to the resolved photon component in the forward
direction\footnote{Since $\eta$ here is defined in the hadronic
centre-of-mass frame, the forward region in the laboratory frame
corresponds to $\eta > -1$.}.  In all $Q^2$ regions, the measured
cross section increases with $\eta^F$ in the region $-2.5< \eta^F <
-1.5$.  For $\eta^F>-1.5$, the cross section decreases as $\eta^F$
increases for $Q^2\gtrsim 10$~GeV$^2$, whereas in photoproduction
and at low $Q^2$ the cross section increases. The NLO
prediction for photoproduction describes the measured cross
section. At low $Q^2$ the NLO prediction using $\mu^2=\q2+\etjetd$
underestimates the measured cross section in the forward direction.
The prediction with $\mu^2 = Q^2$ again agrees reasonably well with
the data within large theoretical uncertainties (not shown).

The differences between the data and NLO calculations may be due to
the persistence of a resolved component at $Q^2>1 \gev^2$. To study this
in more detail, the ratio of dijet cross sections for high and low
$\xgo$ values is presented in the next subsection.

\subsection{Ratios of dijet cross sections}

The $Q^2$ dependence of the direct- and resolved-enhanced components
of the dijet cross section has been studied in more detail using the
ratio
\begin{equation*}
R =\frac{\sigma (\xgo < 0.75) }{ \sigma (\xgo > 0.75)}\;.
\end{equation*}
A number of experimental and theoretical uncertainties cancel in this ratio,
so that the presence of a resolved contribution can be investigated at
higher precision than in the individual cross sections.

\begin{figure}
\centerline{\epsfig{file=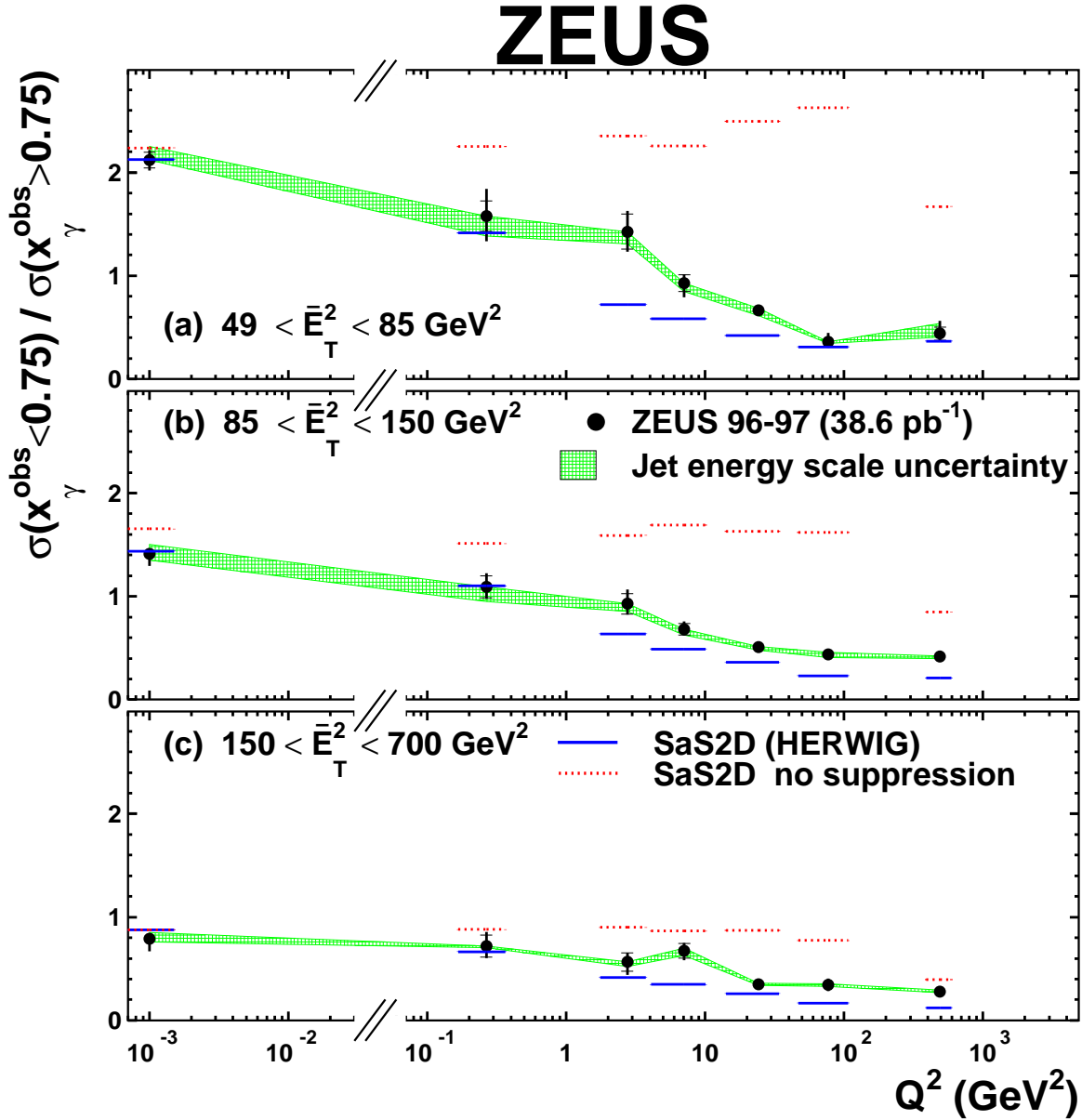,height=17.0cm,clip=}}
\caption{
Measured ratio 
$R =\sigma (\xgo < 0.75) /\sigma (\xgo > 0.75)$ 
as a function of $Q^2$ in different regions of $\et2$ (black dots).
The LO+PS calculations of {\sc Herwig} using the SaS2D photon PDFs are also shown.
Other details are as in the caption to Fig.~1.}
\label{virtualgamma_fig5}
\end{figure}

\begin{figure}
\centerline{\epsfig{file=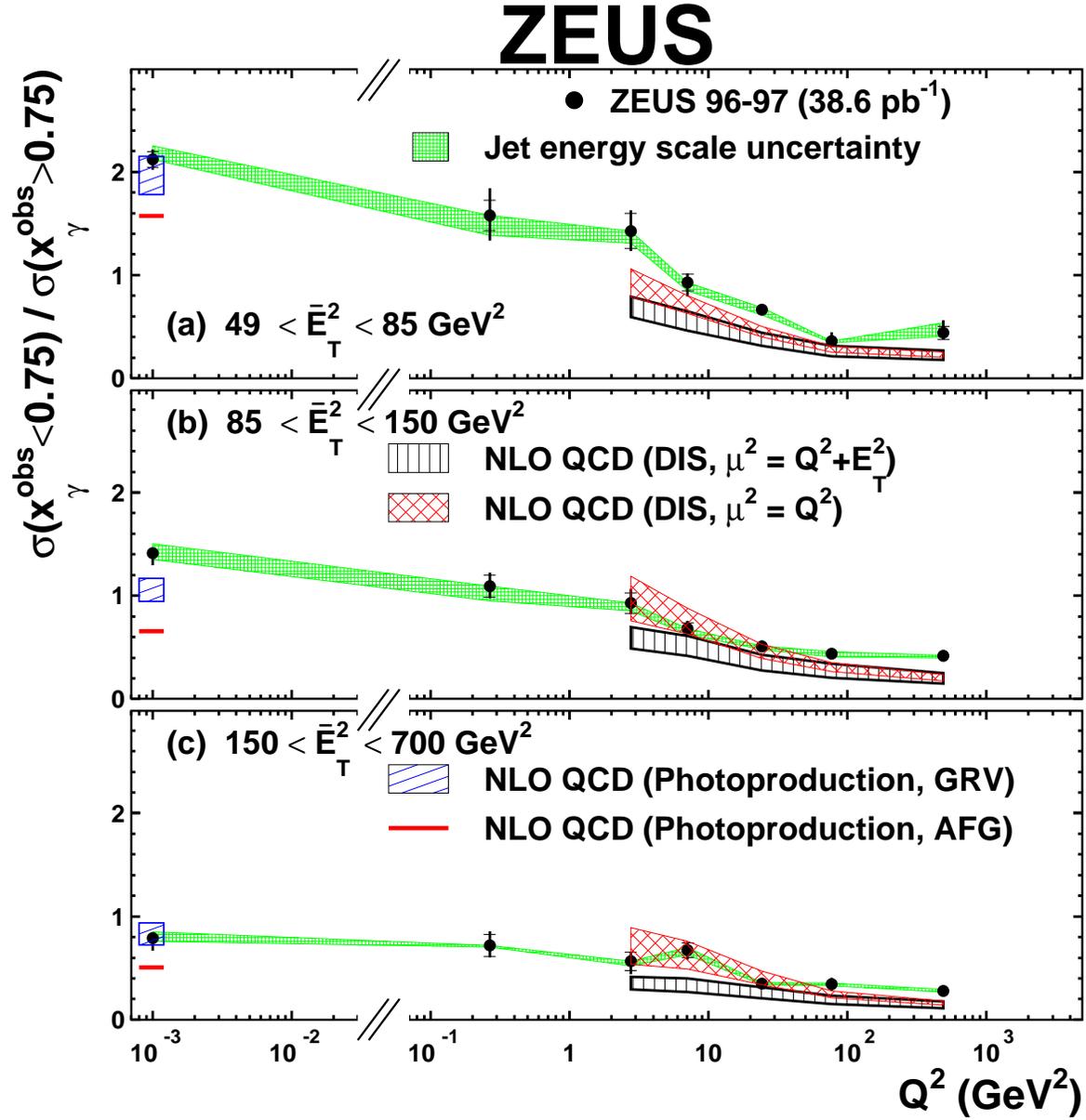,height=17.0cm,clip=}}
\caption{
Measured ratio 
$R =\sigma (\xgo < 0.75) /\sigma (\xgo > 0.75)$ 
as a function of $Q^2$ in different regions of $\et2$ (black dots).
The NLO QCD calculations of {\sc Disaster++} with  $\mu^2=\q2+\etjetd$
and $\mu^2=\q2$ as well as the Frixione and Ridolfi predictions for
the photoproduction region are also shown. The hatched bands represent the
theoretical uncertainties.
Other details are as in the caption to Fig.~1.}
\label{virtualgamma_fig6}
\end{figure}

Figures~\ref{virtualgamma_fig5} and~\ref{virtualgamma_fig6} show the
ratio $R$ as a function of $Q^2$ in three different regions of $\et2$.
The $Q^2$ dependence of the data is stronger at low $\et2$ than at
higher $\et2$, showing that the low-$\xgo$ component is suppressed at
low $Q^2$ as $\et2$ increases, and at low $\et2$ as $Q^2$ increases.
The ratio is also given in Table~\ref{table-ratio}.

Predictions of the {\sc Herwig} MC program using the SaS2D
parameterisation of the photon PDFs are compared to the data in
Fig.~\ref{virtualgamma_fig5}. The SaS2D parameterisation contains the
suppression of the virtual photon structure with increasing $Q^2$. The
predictions fall with increasing $Q^2$ and qualitatively reproduce the
data.  However, the predictions using SaS2D with the
suppression of the virtual photon structure switched off
are relatively constant with $Q^2$.

The NLO calculations are compared to the data in
Fig.~\ref{virtualgamma_fig6}.  The photoproduction calculations using
GRV are in reasonable agreement with the data, whereas those using AFG
are below the data.  In the DIS region, the predictions lie below the
data at low $\et2$.  However, some suppression in the ratio as a
function of $Q^2$ is observed.

\section{Summary and conclusions}

Dijet differential cross sections have been measured in the range $0
\lesssim Q^2 < 2000$~GeV$^2$ with $0.2<y<0.55$, $-3<\eta^{\rm jet}<0$
and $E_T^{{\rm jet1}, {\rm jet2}} > 7.5$ and $6.5 $~GeV, as a function
of $Q^2$, $E_T^{\rm jet1}$ and $\eta^F$ in the photon-proton
centre-of-mass frame.  These precise measurements, spanning a large
range of photon virtualities, including photoproduction, DIS, and the
transition region between them, are qualitatively described by
leading-logarithmic parton-shower MC models which introduce virtual
photon structure, suppressed with increasing $Q^2$.
These data may constrain such parton densities significantly if used
in future fits.

The currently available next-to-leading-order QCD calculations have
large uncertainties at low $\q2$, where the presence of a resolved-photon
contribution may be expected. Improved higher-order or resummed
calculations are needed. In DIS, the NLO QCD predictions generally
underestimate the cross section at low-$\xgo$ relative to that at high
$\xgo$.

\vspace{0.5cm}
\noindent {\Large\bf Acknowledgements}
\vspace{0.3cm}

The design, construction and installation of the ZEUS detector have
been made possible by the ingenuity and dedicated efforts of many
people who are not listed as authors. Their contributions are
acknowledged with great appreciation. The experiment was made possible
by the inventiveness and the diligent efforts of the HERA machine
group. The strong support and encouragement of the DESY directorate
have been invaluable.

\providecommand{\etal}{et al.\xspace}
\providecommand{\coll}{Coll.\xspace}

\begin{table}
 \begin{center}
   \begin{tabular}{|c|ccccc|} \hline
    $Q^2$ {\rm bin (GeV$^2$)} &  $d\sigma/dQ^2$ & $\Delta_{\rm stat}$ & $\Delta_{\rm syst}$ & $\Delta_{\rm ES}$ & (pb/GeV$^2$) \\ \hline \hline
       0, 1 & 9280  & $\pm 113$  & $^{+102}_{-69.3}$  & $_{-1100}^{+917}$  & \\ \hline
       0.1, 0.55 & 2250  & $\pm 45.2$  & $^{+215}_{-210}$  & $_{-227}^{+194}$  & \\ \hline
       1.5, 4.5 & 167  & $\pm 2.22$  & $^{+14.1}_{-11.9}$  & $_{-15.2}^{+12.5}$  & \\ \hline
       4.5, 10.5 & 54.5  & $\pm 0.54$  & $^{+3.05}_{-2.76}$  & $_{-4.48}^{+3.94}$  & \\ \hline
       10.5, 49 & 11.9  & $\pm 0.093$  & $^{+0.79}_{-0.32}$  & $_{-1.01}^{+0.81}$  & \\ \hline
       49, 120 & 2.27  & $\pm 0.027$  & $^{+0.14}_{-0.13}$  & $_{-0.2}^{+0.17}$  & \\ \hline
       120, 2000 & 0.095  & $\pm 0.0011$  & $^{+0.0019}_{-0.0048}$  & $_{-0.0068}^{+0.0059}$  & \\ \hline
   \end{tabular}
 \end{center}
\caption[]{
Measured dijet cross-sections $d \sigma / d Q^2$.
The statistical, systematic and jet energy scale, $\Delta_{\rm ES}$, uncertainties are shown 
separately.
}\label{table-dsdq2}
\end{table}

\begin{table}
 \begin{center}
   \begin{tabular}{|c|cccc|} \hline
    $Q^2$ {\rm bin (GeV$^2$)} &  $d\sigma/dQ^2$ & $\Delta_{\rm stat}$ & $\Delta_{\rm syst}$ & (pb/GeV$^2$) \\ \hline \hline
       0, 1 & 5710  & $\pm 96.4$  & $^{+63}_{-41.7}$  & \\ \hline
       0.1, 0.55 & 1270  & $\pm 32.3$  & $^{+197}_{-141}$  & \\ \hline
       1.5, 4.5 & 87.8  & $\pm 1.54$  & $^{+8.89}_{-7.01}$  & \\ \hline
       4.5, 10.5 & 24.5  & $\pm 0.35$  & $^{+1.39}_{-2.22}$  & \\ \hline
       10.5, 49 & 4.21  & $\pm 0.051$  & $^{+0.53}_{-0.27}$  & \\ \hline
       49, 120 & 0.72  & $\pm 0.015$  & $^{+0.17}_{-0.1}$  & \\ \hline
       120, 2000 & 0.022  & $\pm 0.00052$  & $^{+0.0081}_{-0.0049}$  & \\ \hline
   \end{tabular}
 \end{center}
\caption[]{
Measured dijet cross-sections $d \sigma / d Q^2$ for $\xgo < 0.75$.
The statistical and systematic uncertainties are shown separately.
}\label{table-dsdq2-lox}
\end{table}

\begin{table}
 \begin{center}
   \begin{tabular}{|c|cccc|} \hline
    $Q^2$ {\rm bin (GeV$^2$)} &  $d\sigma/dQ^2$ & $\Delta_{\rm stat}$ & $\Delta_{\rm syst}$ & (pb/GeV$^2$) \\ \hline \hline
       0, 1 & 3620  & $\pm 65.8$  & $^{+39.2}_{-31.3}$  & \\ \hline
       0.1, 0.55 & 980  & $\pm 31.9$  & $^{+32.8}_{-87.2}$  & \\ \hline
       1.5, 4.5 & 79.5  & $\pm 1.61$  & $^{+6.01}_{-4.06}$  & \\ \hline
       4.5, 10.5 & 30  & $\pm 0.42$  & $^{+2.03}_{-0.63}$  & \\ \hline
       10.5, 49 & 7.74  & $\pm 0.08$  & $^{+0.5}_{-0.27}$  & \\ \hline
       49, 120 & 1.63  & $\pm 0.024$  & $^{+0.15}_{-0.13}$  & \\ \hline
       120, 2000 & 0.077  & $\pm 0.0011$  & $^{+0.0013}_{-0.0057}$  & \\ \hline
   \end{tabular}
 \end{center}
\caption[]{
Measured dijet cross-sections $d \sigma / d Q^2$ for $\xgo > 0.75$.
The statistical and systematic uncertainties are shown separately.
}\label{table-dsdq2-hix}
\end{table}


\begin{table}
 \begin{center}
   \begin{tabular}{|c|ccccc|} \hline
    $E_{T}^{\rm jet1}$ bin (GeV) &  $d^2\sigma/dE_{T}^{\rm jet1} dQ^2$ & $\Delta_{\rm stat}$ & $\Delta_{\rm syst}$ & $\Delta_{\rm ES}$ & (pb/GeV$^3$) \\ \hline \hline
      \multicolumn{6}{|c|}{$0 < Q^2 < 1$~GeV$^2$} \\ \hline
       7.5, 10 & 1740  & $\pm 34.8$  & $^{+20.9}_{-84.3}$  & $_{-189}^{+146}$  & \\ \hline
       10, 13 & 1010  & $\pm 20.4$  & $^{+11.3}_{-24.8}$  & $_{-118}^{+90}$  & \\ \hline
       13, 17 & 337  & $\pm 8.5$  & $^{+21.5}_{-8.51}$  & $_{-39.9}^{+35.7}$  & \\ \hline
       17, 22 & 76.4  & $\pm 3.22$  & $^{+3.9}_{-3.14}$  & $_{-9.6}^{+9.45}$  & \\ \hline
       22, 29 & 15  & $\pm 1.25$  & $^{+0.1}_{-0.085}$  & $_{-2.03}^{+1.96}$  & \\ \hline
       29, 50 & 1.75  & $\pm 0.092$  & $^{+0.37}_{-0.16}$  & $_{-0.23}^{+0.22}$  & \\ \hline
      \multicolumn{6}{|c|}{$0.1 < Q^2 < 0.55$~GeV$^2$} \\ \hline
       7.5, 10 & 484  & $\pm 22.6$  & $^{+24.3}_{-75.5}$  & $_{-70.3}^{+53.6}$  & \\ \hline
       10, 13 & 243  & $\pm 12.1$  & $^{+6.81}_{-14.7}$  & $_{-15}^{+16.8}$  & \\ \hline
       13, 17 & 68.7  & $\pm 4.91$  & $^{+4.92}_{-1.38}$  & $_{-3.53}^{+4.22}$  & \\ \hline
       17, 22 & 17.9  & $\pm 2.19$  & $^{+1.33}_{-1.76}$  & $_{-0.94}^{+0.48}$  & \\ \hline
       22, 29 & 2.49  & $\pm 0.67$  & $^{+0.39}_{-0.29}$  & $_{-0.69}^{+0.13}$  & \\ \hline
       29, 50 & 0.38  & $\pm 0.17$  & $^{+0.071}_{-0.083}$  & $_{-0}^{+0.048}$  & \\ \hline
      \multicolumn{6}{|c|}{$1.5 < Q^2 < 4.5$~GeV$^2$} \\ \hline
       7.5, 10 & 35.2  & $\pm 0.72$  & $^{+2.46}_{-1.31}$  & $_{-4.71}^{+3.44}$  & \\ \hline
       10, 13 & 17.6  & $\pm 0.41$  & $^{+0.96}_{-1}$  & $_{-1.12}^{+0.96}$  & \\ \hline
       13, 17 & 5.09  & $\pm 0.16$  & $^{+0.52}_{-0.044}$  & $_{-0.26}^{+0.36}$  & \\ \hline
       17, 22 & 1.64  & $\pm 0.091$  & $^{+0.032}_{-0.13}$  & $_{-0.086}^{+0.057}$  & \\ \hline
       22, 29 & 0.32  & $\pm 0.033$  & $^{+0.0087}_{-0.035}$  & $_{-0.01}^{+0.015}$  & \\ \hline
       29, 50 & 0.037  & $\pm 0.0063$  & $^{+0.0023}_{-0.0049}$  & $_{-0.0035}^{+0.0024}$  & \\ \hline
      \multicolumn{6}{|c|}{$4.5 < Q^2 < 10.5$~GeV$^2$} \\ \hline
       7.5, 10 & 10.2  & $\pm 0.16$  & $^{+0.23}_{-1.32}$  & $_{-1.21}^{+1}$  & \\ \hline
       10, 13 & 5.86  & $\pm 0.1$  & $^{+0.25}_{-0.16}$  & $_{-0.37}^{+0.34}$  & \\ \hline
       13, 17 & 2.08  & $\pm 0.048$  & $^{+0.043}_{-0.061}$  & $_{-0.12}^{+0.11}$  & \\ \hline
       17, 22 & 0.52  & $\pm 0.02$  & $^{+0.0084}_{-0.033}$  & $_{-0.019}^{+0.034}$  & \\ \hline
       22, 29 & 0.13  & $\pm 0.0097$  & $^{+0.0071}_{-0.0078}$  & $_{-0.011}^{+0.0053}$  & \\ \hline
       29, 50 & 0.011  & $\pm 0.0015$  & $^{+0.002}_{-0.00084}$  & $_{-0.00059}^{+0.00039}$  & \\ \hline
   \end{tabular}
 \end{center}
\caption[]{
Measured dijet cross-section $d^2 \sigma / d Q^2 d E_T^{\rm jet1}$.
}\label{table-q2etl-1}
\end{table}
\begin{table}
 \begin{center}
   \begin{tabular}{|c|ccccc|} \hline
    $E_{T}^{\rm jet1}$ bin (GeV) &  $d^2\sigma/dE_{T}^{\rm jet1} dQ^2$ & $\Delta_{\rm stat}$ & $\Delta_{\rm syst}$ & $\Delta_{\rm ES}$ & (pb/GeV$^3$) \\ \hline \hline
      \multicolumn{6}{|c|}{$10.5 < Q^2 < 49$~GeV$^2$} \\ \hline
       7.5, 10 & 1.9  & $\pm 0.024$  & $^{+0.15}_{-0.059}$  & $_{-0.25}^{+0.17}$  & \\ \hline
       10, 13 & 1.36  & $\pm 0.018$  & $^{+0.057}_{-0.041}$  & $_{-0.082}^{+0.091}$  & \\ \hline
       13, 17 & 0.51  & $\pm 0.0091$  & $^{+0.024}_{-0.013}$  & $_{-0.025}^{+0.028}$  & \\ \hline
       17, 22 & 0.14  & $\pm 0.0041$  & $^{+0.0029}_{-0.007}$  & $_{-0.0086}^{+0.0031}$  & \\ \hline
       22, 29 & 0.036  & $\pm 0.0018$  & $^{+0.0014}_{-0.0021}$  & $_{-0.0012}^{+0.0017}$  & \\ \hline
       29, 50 & 0.0031  & $\pm 0.00031$  & $^{+0.00018}_{-0.00011}$  & $_{-0.00034}^{+0.00028}$  & \\ \hline
      \multicolumn{6}{|c|}{$49 < Q^2 < 120$~GeV$^2$} \\ \hline
       7.5, 10 & 0.34  & $\pm 0.0071$  & $^{+0.025}_{-0.024}$  & $_{-0.053}^{+0.038}$  & \\ \hline
       10, 13 & 0.23  & $\pm 0.0049$  & $^{+0.014}_{-0.011}$  & $_{-0.016}^{+0.018}$  & \\ \hline
       13, 17 & 0.11  & $\pm 0.0031$  & $^{+0.0016}_{-0.0071}$  & $_{-0.0047}^{+0.0058}$  & \\ \hline
       17, 22 & 0.039  & $\pm 0.0016$  & $^{+0.00087}_{-0.001}$  & $_{-0.0014}^{+0.0011}$  & \\ \hline
       22, 29 & 0.0059  & $\pm 0.00047$  & $^{+0.001}_{-0.00023}$  & $_{-0.00018}^{+0.00031}$  & \\ \hline
       29, 50 & 0.00072  & $\pm 8.5\times 10^{-5}$  & $^{+0.00023}_{-3\times 10^{-5}}$  & $_{-5.2\times 10^{-5}}^{+8.3\times 10^{-5}}$  & \\ \hline
      \multicolumn{6}{|c|}{$120 < Q^2 < 2000$~GeV$^2$} \\ \hline
       7.5, 10 & 0.01  & $\pm 0.00025$  & $^{+0.00041}_{-0.00098}$  & $_{-0.0014}^{+0.001}$  & \\ \hline
       10, 13 & 0.0099  & $\pm 0.00021$  & $^{+0.00043}_{-0.00076}$  & $_{-0.0006}^{+0.00059}$  & \\ \hline
       13, 17 & 0.0056  & $\pm 0.00014$  & $^{+0.00011}_{-0.00023}$  & $_{-0.00025}^{+0.00022}$  & \\ \hline
       17, 22 & 0.0021  & $\pm 7.7\times 10^{-5}$  & $^{+6.7\times 10^{-5}}_{-0.00011}$  & $_{-6.4\times 10^{-5}}^{+0.00014}$  & \\ \hline
       22, 29 & 0.00069  & $\pm 3.6\times 10^{-5}$  & $^{+7.1\times 10^{-5}}_{-3.9\times 10^{-5}}$  & $_{-3.3\times 10^{-5}}^{+1.8\times 10^{-5}}$  & \\ \hline
       29, 50 & $6\times 10^{-5}$  & $\pm 5.2\times 10^{-6}$  & $^{+5\times 10^{-6}}_{-4.2\times 10^{-6}}$  & $_{-2\times 10^{-6}}^{+3.4\times 10^{-6}}$  & \\ \hline
   \end{tabular}
 \end{center}
\caption[]{
Measured dijet cross-section $d^2 \sigma / d Q^2 d E_T^{\rm jet1}$.
}\label{table-q2etl-2}
\end{table}
\newpage

\begin{table}
 \begin{center}
   \begin{tabular}{|c|ccccc|} \hline
    $\eta^F$ bin &  $d^2\sigma / d\eta^F dQ^2$ & $\Delta_{\rm stat}$ & $\Delta_{\rm syst}$ & $\Delta_{\rm ES}$ & (pb/GeV$^2$) \\ \hline \hline
      \multicolumn{6}{|c|}{$0 < Q^2 < 1$~GeV$^2$} \\ \hline
       -3, -1.8 & 771  & $\pm 27.5$  & $^{+39.1}_{-38.3}$  & $_{-119}^{+92.8}$  & \\ \hline
       -1.8, -1.4 & 3790  & $\pm 108$  & $^{+130}_{-49.8}$  & $_{-335}^{+305}$  & \\ \hline
       -1.4, -0.8 & 4510  & $\pm 96.9$  & $^{+162}_{-55.9}$  & $_{-571}^{+399}$  & \\ \hline
       -0.8, 0 & 5300  & $\pm 99.3$  & $^{+66.6}_{-229}$  & $_{-606}^{+494}$  & \\ \hline
      \multicolumn{6}{|c|}{$0.1 < Q^2 < 0.55$~GeV$^2$} \\ \hline
       -3, -1.8 & 229  & $\pm 18.1$  & $^{+3.41}_{-28.6}$  & $_{-21.3}^{+35.3}$  & \\ \hline
       -1.8, -1.4 & 939  & $\pm 64.2$  & $^{+77.3}_{-67.3}$  & $_{-98.9}^{+60.9}$  & \\ \hline
       -1.4, -0.8 & 1070  & $\pm 55$  & $^{+132}_{-81.6}$  & $_{-102}^{+82.2}$  & \\ \hline
       -0.8, 0 & 1240  & $\pm 58.7$  & $^{+165}_{-155}$  & $_{-126}^{+101}$  & \\ \hline
      \multicolumn{6}{|c|}{$1.5 < Q^2 < 4.5$~GeV$^2$} \\ \hline
       -3, -1.8 & 15.3  & $\pm 0.62$  & $^{+1.62}_{-2.08}$  & $_{-2.05}^{+1.68}$  & \\ \hline
       -1.8, -1.4 & 75.8  & $\pm 2.47$  & $^{+4.11}_{-8.29}$  & $_{-5.4}^{+5.58}$  & \\ \hline
       -1.4, -0.8 & 81.5  & $\pm 1.96$  & $^{+10}_{-1.15}$  & $_{-7.21}^{+5.05}$  & \\ \hline
       -0.8, 0 & 87.1  & $\pm 1.78$  & $^{+7.4}_{-8.2}$  & $_{-7.57}^{+6.37}$  & \\ \hline
      \multicolumn{6}{|c|}{$4.5 < Q^2 < 10.5$~GeV$^2$} \\ \hline
       -3, -1.8 & 6.67  & $\pm 0.19$  & $^{+0.37}_{-0.62}$  & $_{-0.85}^{+0.78}$  & \\ \hline
       -1.8, -1.4 & 26  & $\pm 0.61$  & $^{+3.17}_{-0.21}$  & $_{-2.06}^{+1.84}$  & \\ \hline
       -1.4, -0.8 & 26.8  & $\pm 0.47$  & $^{+0.27}_{-2.12}$  & $_{-1.81}^{+1.63}$  & \\ \hline
       -0.8, 0 & 24.9  & $\pm 0.41$  & $^{+2.16}_{-1.58}$  & $_{-1.97}^{+1.62}$  & \\ \hline
   \end{tabular}
 \end{center}
\caption[]{
Measured dijet cross-section $d^2 \sigma / d Q^2 d  \eta^F$.
}\label{table-q2eta-1}
\end{table}
\begin{table}
 \begin{center}
   \begin{tabular}{|c|ccccc|} \hline
    $\eta^F$ bin &  $d^2\sigma / d\eta^F dQ^2$ & $\Delta_{\rm stat}$ & $\Delta_{\rm syst}$ & $\Delta_{\rm ES}$ & (pb/GeV$^2$) \\ \hline \hline
      \multicolumn{6}{|c|}{$10.5 < Q^2 < 49$~GeV$^2$} \\ \hline
       -3, -1.8 & 1.45  & $\pm 0.031$  & $^{+0.17}_{-0.036}$  & $_{-0.2}^{+0.15}$  & \\ \hline
       -1.8, -1.4 & 6.57  & $\pm 0.11$  & $^{+0.48}_{-0.25}$  & $_{-0.47}^{+0.44}$  & \\ \hline
       -1.4, -0.8 & 6.21  & $\pm 0.084$  & $^{+0.37}_{-0.16}$  & $_{-0.45}^{+0.36}$  & \\ \hline
       -0.8, 0 & 4.78  & $\pm 0.064$  & $^{+0.4}_{-0.22}$  & $_{-0.37}^{+0.28}$  & \\ \hline
      \multicolumn{6}{|c|}{$49 < Q^2 < 120$~GeV$^2$} \\ \hline
       -3, -1.8 & 0.3  & $\pm 0.01$  & $^{+0.032}_{-0.042}$  & $_{-0.045}^{+0.036}$  & \\ \hline
       -1.8, -1.4 & 1.27  & $\pm 0.034$  & $^{+0.14}_{-0.08}$  & $_{-0.1}^{+0.098}$  & \\ \hline
       -1.4, -0.8 & 1.24  & $\pm 0.026$  & $^{+0.074}_{-0.08}$  & $_{-0.09}^{+0.076}$  & \\ \hline
       -0.8, 0 & 0.87  & $\pm 0.018$  & $^{+0.065}_{-0.056}$  & $_{-0.065}^{+0.059}$  & \\ \hline
      \multicolumn{6}{|c|}{$120 < Q^2 < 2000$~GeV$^2$} \\ \hline
       -3, -1.8 & 0.0097  & $\pm 0.00035$  & $^{+9.2\times 10^{-5}}_{-0.0011}$  & $_{-0.0014}^{+0.00093}$  & \\ \hline
       -1.8, -1.4 & 0.055  & $\pm 0.0015$  & $^{+0.0011}_{-0.0067}$  & $_{-0.0032}^{+0.0035}$  & \\ \hline
       -1.4, -0.8 & 0.051  & $\pm 0.001$  & $^{+0.003}_{-0.0012}$  & $_{-0.0032}^{+0.0027}$  & \\ \hline
       -0.8, 0 & 0.039  & $\pm 0.00082$  & $^{+0.002}_{-0.0019}$  & $_{-0.0024}^{+0.0022}$  & \\ \hline
   \end{tabular}
 \end{center}
\caption[]{
Measured dijet cross-section $d^2 \sigma / d Q^2 d  \eta^F$.
}\label{table-q2eta-2}
\end{table}
\newpage

\begin{table}
 \begin{center}
   \begin{tabular}{|c|ccccc|} \hline
    $Q^2$  bin &  $R$ & $\Delta_{\rm stat}$ & $\Delta_{\rm syst}$ & $\Delta_{\rm ES}$ &  \\ \hline \hline
      \multicolumn{6}{|c|}{$49 < \et2 < 85$~GeV$^2$} \\ \hline
       0, 1 & 2.12  & $\pm 0.075$  & $^{+0.057}_{-0.066}$  & $_{-0.12}^{+0.0026}$  & \\ \hline
       0.1, 0.55 & 1.57  & $\pm 0.14$  & $^{+0.22}_{-0.19}$  & $_{-0}^{+0.19}$  & \\ \hline
       1.5, 4.5 & 1.42  & $\pm 0.16$  & $^{+0.1}_{-0.096}$  & $_{-0}^{+0.11}$  & \\ \hline
       4.5, 10.5 & 0.92  & $\pm 0.08$  & $^{+0.035}_{-0.1}$  & $_{-0}^{+0.078}$  & \\ \hline
       10.5, 49 & 0.66  & $\pm 0.039$  & $^{+0.029}_{-0.034}$  & $_{-0.021}^{+0.049}$  & \\ \hline
       49, 120 & 0.35  & $\pm 0.037$  & $^{+0.079}_{-0.018}$  & $_{-0.0021}^{+0.014}$  & \\ \hline
       120, 2000 & 0.44  & $\pm 0.063$  & $^{+0.1}_{-0.011}$  & $_{-0.1}^{+0.035}$  & \\ \hline
      \multicolumn{6}{|c|}{$85 < \et2 < 150$~GeV$^2$} \\ \hline
       0, 1 & 1.41  & $\pm 0.048$  & $^{+0.016}_{-0.1}$  & $_{-0.086}^{+0.065}$  & \\ \hline
       0.1, 0.55 & 1.09  & $\pm 0.1$  & $^{+0.079}_{-0.044}$  & $_{-0}^{+0.13}$  & \\ \hline
       1.5, 4.5 & 0.92  & $\pm 0.1$  & $^{+0.095}_{-0.032}$  & $_{-0}^{+0.073}$  & \\ \hline
       4.5, 10.5 & 0.68  & $\pm 0.057$  & $^{+0.052}_{-0.026}$  & $_{-0}^{+0.053}$  & \\ \hline
       10.5, 49 & 0.51  & $\pm 0.027$  & $^{+0.044}_{-0.023}$  & $_{-0.0036}^{+0.033}$  & \\ \hline
       49, 120 & 0.43  & $\pm 0.038$  & $^{+0.029}_{-0.015}$  & $_{-0.019}^{+0.029}$  & \\ \hline
       120, 2000 & 0.41  & $\pm 0.048$  & $^{+0.033}_{-0.0084}$  & $_{-0.0062}^{+0.018}$  & \\ \hline
      \multicolumn{6}{|c|}{$150 < \et2 < 700$~GeV$^2$} \\ \hline
       0, 1 & 0.78  & $\pm 0.032$  & $^{+0.0022}_{-0.11}$  & $_{-0.06}^{+0.028}$  & \\ \hline
       0.1, 0.55 & 0.72  & $\pm 0.1$  & $^{+0.086}_{-0.049}$  & $_{-0}^{+0.018}$  & \\ \hline
       1.5, 4.5 & 0.56  & $\pm 0.088$  & $^{+0.015}_{-0.088}$  & $_{-0}^{+0.042}$  & \\ \hline
       4.5, 10.5 & 0.67  & $\pm 0.07$  & $^{+0.03}_{-0.059}$  & $_{-0.028}^{+0.043}$  & \\ \hline
       10.5, 49 & 0.34  & $\pm 0.026$  & $^{+0.015}_{-0.0057}$  & $_{-0.0098}^{+0.0092}$  & \\ \hline
       49, 120 & 0.34  & $\pm 0.039$  & $^{+0.046}_{-0.049}$  & $_{-0.0068}^{+0.016}$  & \\ \hline
       120, 2000 & 0.28  & $\pm 0.046$  & $^{+0.019}_{-0.0052}$  & $_{-0.011}^{+0.014}$  & \\ \hline
   \end{tabular}
 \end{center}
\caption[]{
Measured ratio $R =\sigma (\xgo < 0.75) /\sigma (\xgo > 0.75)$ 
as a function of $Q^2$ in different regions of $\et2$.
}\label{table-ratio}
\end{table}

\newpage
\clearpage

\newpage
\clearpage

\newpage
\clearpage

\newpage
\clearpage

\newpage
\clearpage

\vfill\eject

\end{document}